\documentclass[11pt]{article}
\usepackage{moriond,epsfig}
\usepackage{xspace}

\usepackage{graphicx}
 \DeclareGraphicsExtensions{.pdf,.jpg,.jpeg,.png}

\def\be{\begin{equation}}
\def\ee{\end{equation}}
\def\bea{\begin{eqnarray}}
\def\eea{\end{eqnarray}}

\newcommand{\cdf}{CDF\xspace}

\newcommand{\dzero}{D\O\xspace}
\newcommand{\ifb}{\ensuremath{\mathrm{fb}^{-1}}\xspace}

\newcommand{\tev}{\ensuremath{\mathrm{\,Te\kern -0.1em V}}\xspace}
\newcommand{\gev}{\ensuremath{\mathrm{\,Ge\kern -0.1em V}}\xspace}
\newcommand{\mev}{\ensuremath{\mathrm{\,Me\kern -0.1em V}}\xspace}
\newcommand{\kev}{\ensuremath{\mathrm{\,ke\kern -0.1em V}}\xspace}
\newcommand{\ev}{\ensuremath{\mathrm{\,e\kern -0.1em V}}\xspace}
\newcommand{\gevc}{\ensuremath{{\mathrm{\,Ge\kern -0.1em V\!/}c}}\xspace}
\newcommand{\mevc}{\ensuremath{{\mathrm{\,Me\kern -0.1em V\!/}c}}\xspace}
\newcommand{\gevcc}{\ensuremath{{\mathrm{\,Ge\kern -0.1em V\!/}c^2}}\xspace}
\newcommand{\mevcc}{\ensuremath{{\mathrm{\,Me\kern -0.1em V\!/}c^2}}\xspace}
\newcommand{\msbottom}{\ensuremath{m_{\tilde{b}}}\xspace}
\newcommand{\mstop}{\ensuremath{m_{\tilde{t}}}\xspace}
\newcommand{\mneutralino}{\ensuremath{m_{\chi^0_1}}\xspace}
\newcommand{\neutralino}{\ensuremath{{\chi^0_1}}\xspace}

\def \et {E_{T}}
\newcommand{\met}{\ensuremath{{{E\!\!\!\!/_T}}}}

\begin{document}
\vspace*{4cm}
\title{Tevatron Searches for New Physics with Photons and Jets}

\author{ Benjamin. P. Brau }

\address{Department of Physics,\\
1126 Lederle Graduate Research Tower,\\
University of Massachusetts,\\
Amherst, MA 0100, USA }

\maketitle\abstracts{
The \dzero and \cdf experiments have each collected more than 8~\ifb in Run II of
Fermilab's Tevatron, and have many recent search results which use up to
5.2~\ifb.
Here I summarize the results of a variety of searches for physics
beyond the Standard Model with an emphasis on searches for very exotic
phenomena.  I will present the status of model-inspired searches for
several signatures of supersymmetry, as well as several other searches
for several ``hidden-valley''  inspired models, all of which contain photons and
jets in the final state.}

\section{Introduction}
Despite its enormous success at explaining the interactions of the
known particles, we are certain that the Standard Model (SM) is incomplete.
We now have overwhelming evidence that the universe contains mostly dark
matter, yet the SM provides no explanation for it.  Among the models on the
market, Supersymmetry (SUSY) is favored by many, in part because it
naturally has a dark matter candidate in the lightest supersymmetric
particle (LSP).  Recently though, a class of models has been proposed that can
generally be described as ``hidden-valley''  (HV) models.  These are characterized
by the general feature that they propose relatively light particles one
might expect to have already discovered, but haven't been because some
mechanism prevents them from interacting with the SM particles, except
gravitationally.  These models therefore also can naturally provide a
solution to the dark matter puzzle.  Most collider experiments sensitive to
new higher energy scales have searched for SUSY in a variety of final
states, and the Tevatron is no exception.  Hidden-valley models are
relatively recent, and searches for them at colliders are just emerging.  I
will discuss several searches for SUSY and HV models at the \cdf
and \dzero experiments at the Tevatron with 2.6-5.2~\ifb.

\section{Searches for Sbottom with $b$ jets and $\met$}
In the MSSM, if $\tan\beta$ is large, there is a large mass splitting in the
third generation.  This leads to relatively low masses for the lightest
scalar bottom quark (sbottom) state.  In an R-parity conserving scenario,
sbottoms are expected to be produced at the Tevatron in pairs via
gluon-gluon fusion or qqbar annihilation.  The sbottom is then kinematically
required to decay into bottom and the lightest neutralino (assumed to be the
LSP), leading to final states which include two b-jets and substantial
missing transverse energy (MET) from the undetected LSP.

Both Tevatron experiments have searched for bbottom in events with jets,
$\met$, and b-tags. The CDF collaboration has searched in 2.65 \ifb of Run
II data\cite{cdfsbottom}, and measures the dominant background in this search from
light-flavor jets misidentified as b-jets with control samples in data.
Events are required to have a good primary vertex, and track activity
consistent with energy measured in the calorimeter.  Events with exactly two
well-measured jets with $E_T > 25$ \gev, not pointing along either jet
direction, and no isolated high-$p_T$ tracks are selected for final
analysis.  The final selection criteria are optimized for two signal regions
in $\Delta M = M_{\tilde{b}_1} - M_{\tilde{\chi}^0_1}$, the mass difference
between the sbottom and the neutralino.  The background predictions for the
two regions are $133.8 \pm 25.2$ events and $47.6 \pm 8.3$ events for the
high and low $\Delta M$ regions respectively.  In the data, 139 events are
observed in the high $\Delta M$ region and 38 in the low $\Delta M$ region.
In the absence of any excess, limits are set in the neutralino-sbottom mass
plane.

\dzero has performed an updated search\cite{d0sbottom}, which uses 5.2 \ifb of Run
II data. This search is also for pair production of scalar bottom
quarks, and is also interpreted in the context of scalar third-generation leptoquarks.
As in the CDF analysis, scalar bottom quarks are assumed to decay to a neutralino
and a $b$ quark.  Events are selected by requiring two or three jets with
$p_T > 20$ \gevc, and no identified isolated electrons or muons.
The $\met$ is required to be greater than 40 \gev, and not aligned with the
direction of any jet.  A neural network $b$-tagging algorithm is employed to
identify heavy-flavor jets,  two of the jets are required to be
consistent with originating from a $b$ quark.  A final selection
on the kinematics of the event, including  the $p_T$ of the leading jet,
$\met$, and the scalar sum of the jets $\et$ and $\met$, $H_T$, is optimized for the smallest expected cross section limit
for each signal mass considered.

\begin{figure}[tb]
\centering
\mbox{\includegraphics[width=7.3cm,clip=]{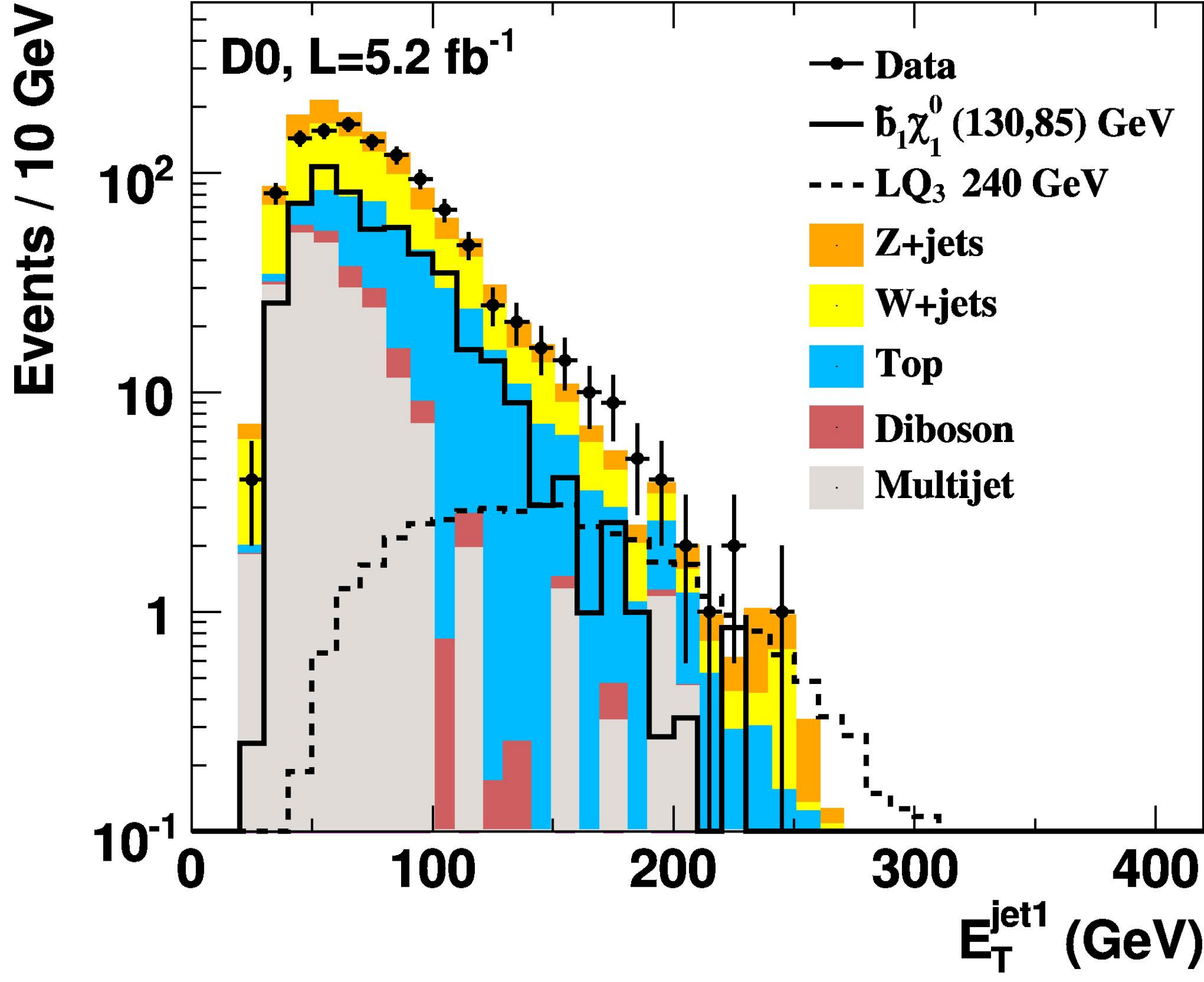}}
\mbox{\includegraphics[width=7.3cm,clip=]{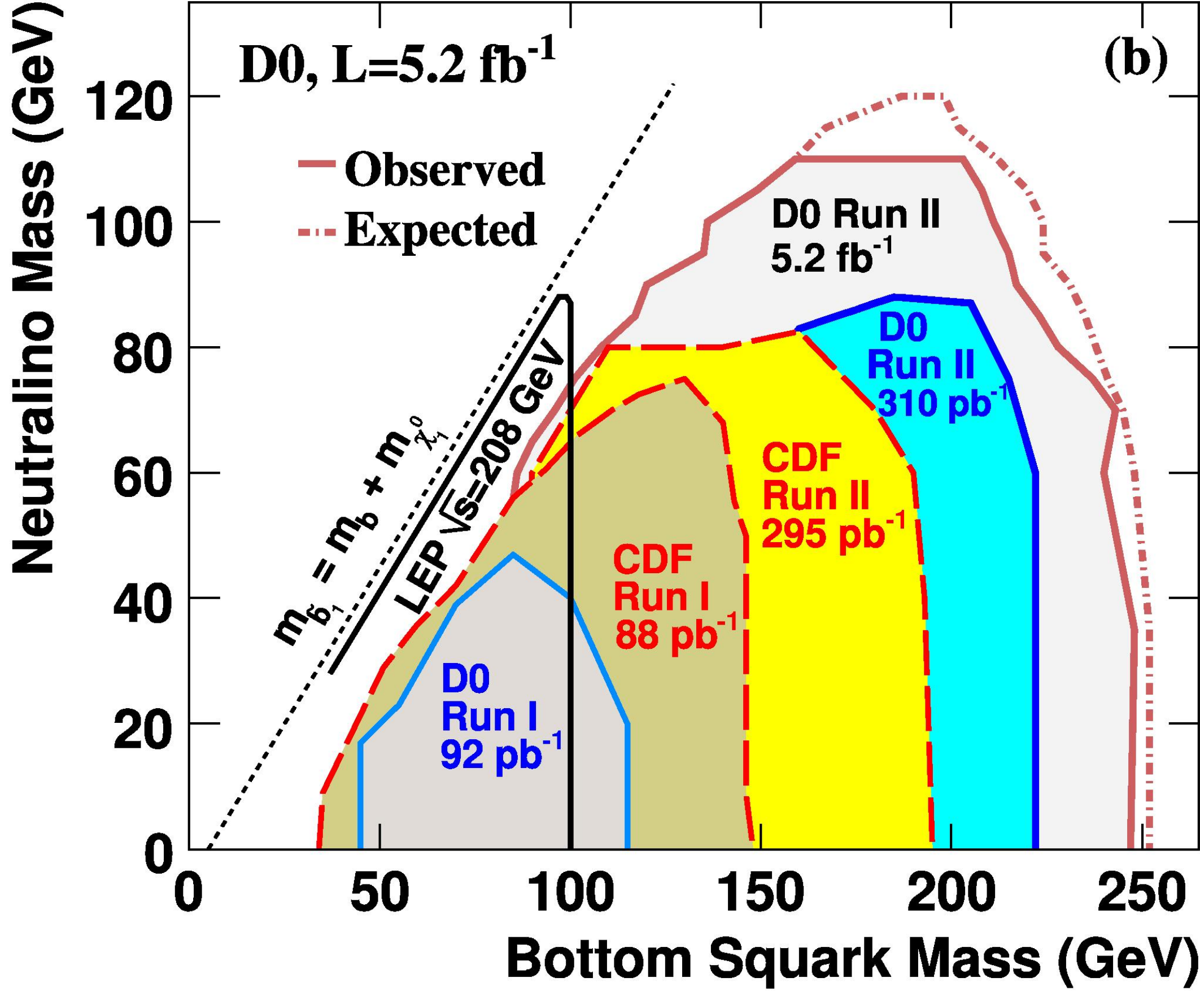}}
\caption{
The $E_T$ of the leading jet (left) and limits in the \mneutralino-\msbottom
neutrino and a $b$ quark plane (right).
\label{fig:sbottom}}
\end{figure}

In the absence of a significant excess, this search has set 95\% C.L. lower
limits on their production in the ( $\msbottom$ , $\mneutralino$ ) mass plane, such
as $\msbottom > 247$ \gevcc for a massless neutralino, and $\mneutralino >$
110 \gevcc for
$160 < \msbottom <200$ \gevcc.
The $E_T$ of the leading jet and limits in the $\mneutralino-\msbottom$
plane are shown in Figure~\ref{fig:sbottom}.
Limits are also placed on third generation
leptoquarks, which are assumed to decay to a tau
neutrino and a $b$ quark.  The 95\% C.L. lower limit on the mass of a
charge-1/3 third-generation scalar leptoquark is 247 \gevcc.

\section{Search for Scalar top with $c$ jets and $\met$}
CDF has also searched for direct stop quark production, where the stop
decays to a charm quark and a neutralino. In this search, the neutralino is
assumed to be the LSP, and R-parity
conservation is assumed. Therefore, the stop signature is two $c$ jets and large
missing transverse energy from the LSP escaping detection.

Events are selected by requiring at least two well-measured jets with $\et > 25 GeV$ and $\met$ $> 50$ GeV.
A neural network is used to reject heavy flavor multi-jet events, and a
second two-dimensional neural network is employed to select charm jets.
The background contribution from heavy-flavor production and light quark
jets misidentified as heavy flavor are estimated from the data.
This search is optimized for a stop mass of 125 \gevcc, and a neutralino
mass of 70 \gevcc.  The predicted background is $132.0\pm 24.4$ events, and
115 events are observed.  For a neutralino mass of 70 \gevcc, and  \mstop
ranging from 115 to 125 \gevcc, the expected signal ranges from 82.4 to 
90.2 events.  No excess is observed and limits are set in the \mneutralino
and \mstop plane, shown in Figure~\ref{fig:stop}.
\begin{figure}[tb]
\centering
\mbox{\includegraphics[width=7.3cm,clip=]{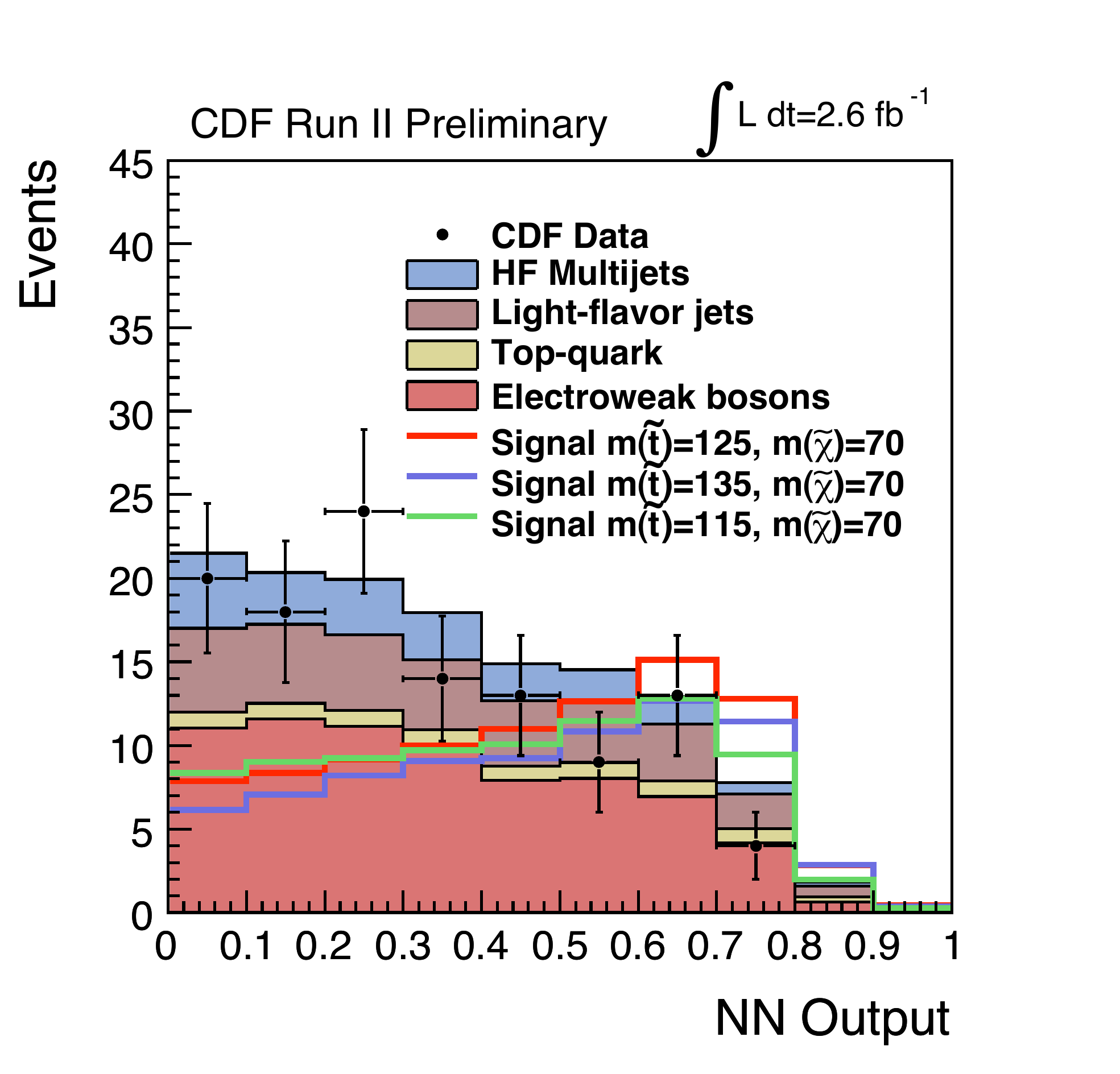}}
\mbox{\includegraphics[width=7.3cm,clip=]{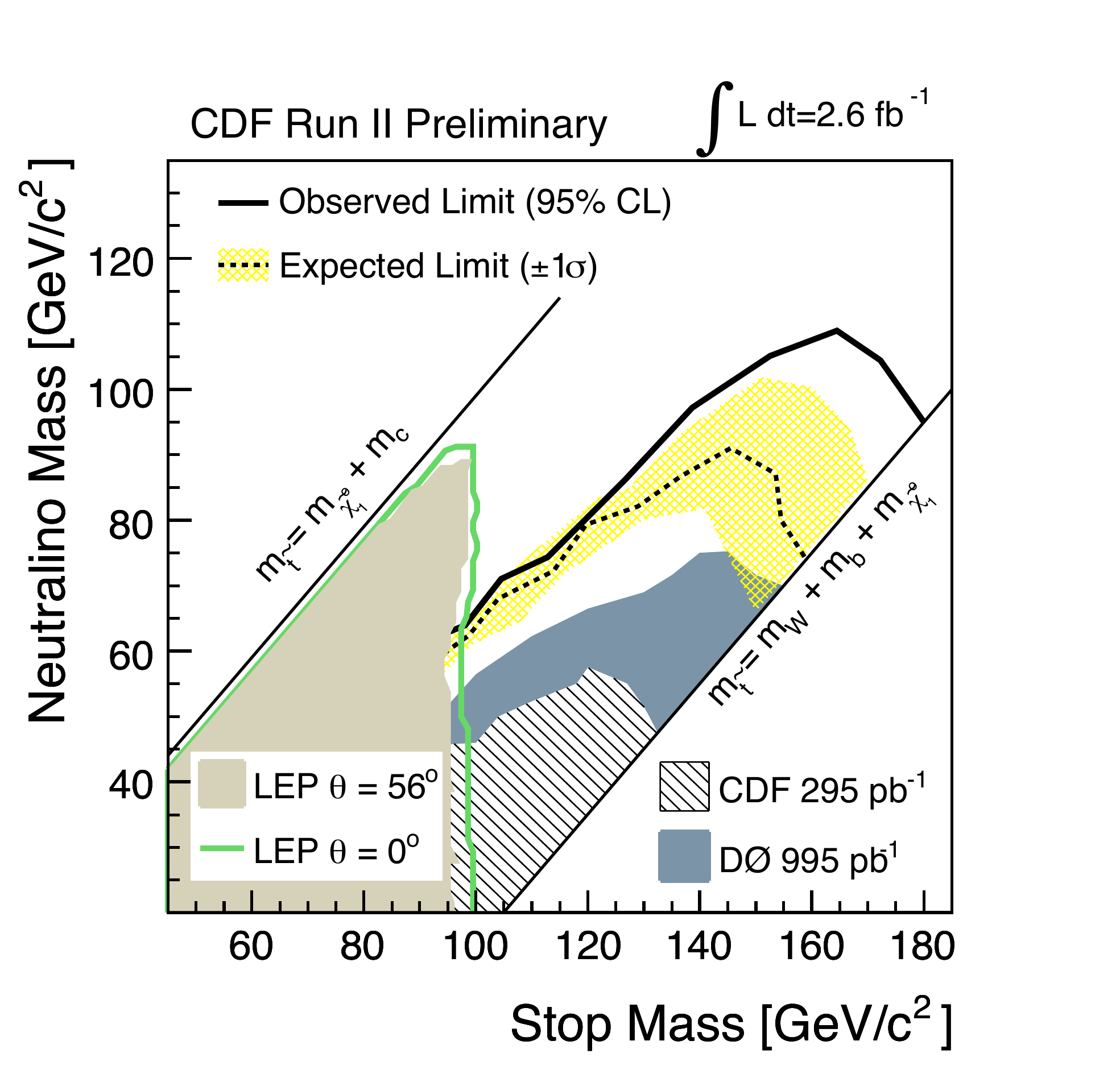}}
\caption{
Output discriminant from the heavy-flavor multi-jet neural network (left) and
the 95\% C.L. limits in the $\mneutralino - \mstop$ mass plane (right).
\label{fig:stop}}
\end{figure}

\section{Search for Gauginos in GMSB}
After the observation in Run 1 by CDF of a two electron, two
photons, and $\met$ candidate event, a class of supersymmetry models with
gauge-mediated SUSY breaking became of particular theoretical interest.
These models solve the ``naturalness problem'' and provide a low-mass dark
matter candidate.  In this search, a scenario where the lightest neutralino,
\neutralino, decays almost exclusively to a photon and a weakly interacting,
stable gravitino is considered.  The gravitino escapes the detector without depositing any
energy, and therefore can generate significant $\met$.  At the Tevatron,
production of gaugino pairs would then result in two photons, $\met$, and
possibly other products of the cascade decay.

CDF has performed a search \cite{Aaltonen:2009tp} for this signature with
2.6 \ifb of Run II data.  This search is optimized for neutralino lifetimes
of order a few ns or less, and utilizes a  timing system allowing the photon
arrival time to be measured.  A detailed model of the $\met$ resolution is
employed to reject backgrounds from instrumental and non-collision sources,
enhancing the sensitivity for large \neutralino masses.  This model enables
a per-event figure of merit to be computed, allowing events where the $\met$
is more likely to be the result of a mis-measured jet or other poorly
reconstructed object to be rejected, while retaining with high efficiency
events with well-measured real $\met$.  

Events are selected by requiring two photons with $\et> 20 \gev$.  The $\met$
significance of events is required to be greater than 3, and the total sum
$\et$ of the observed jets and $\met$ in the event is required to be more than
200 \gev.  Finally, the two photons are required to not be back-to-back by
requiring the angle between them to be less than $\pi - 0.35$ radians.
After all selection, the dominant background originates from electroweak
production of a $Z^0$ with two photons, where the $Z^0$ decays to neutrinos
and results in significant $\met$.  The total predicted background in this
search is $1.4  \pm 0.4$ events, and zero are observed.  Limits are set at
95\% C.L. in the \mneutralino and \neutralino lifetime plane, shown in Figure~\ref{fig:ggmet}.  For a
short-lived neutralino with lifetime less than a nanosecond, the limit is
150 \gevcc.

\begin{figure}[tb]
\centering
\mbox{\includegraphics[width=7.3cm,clip=]{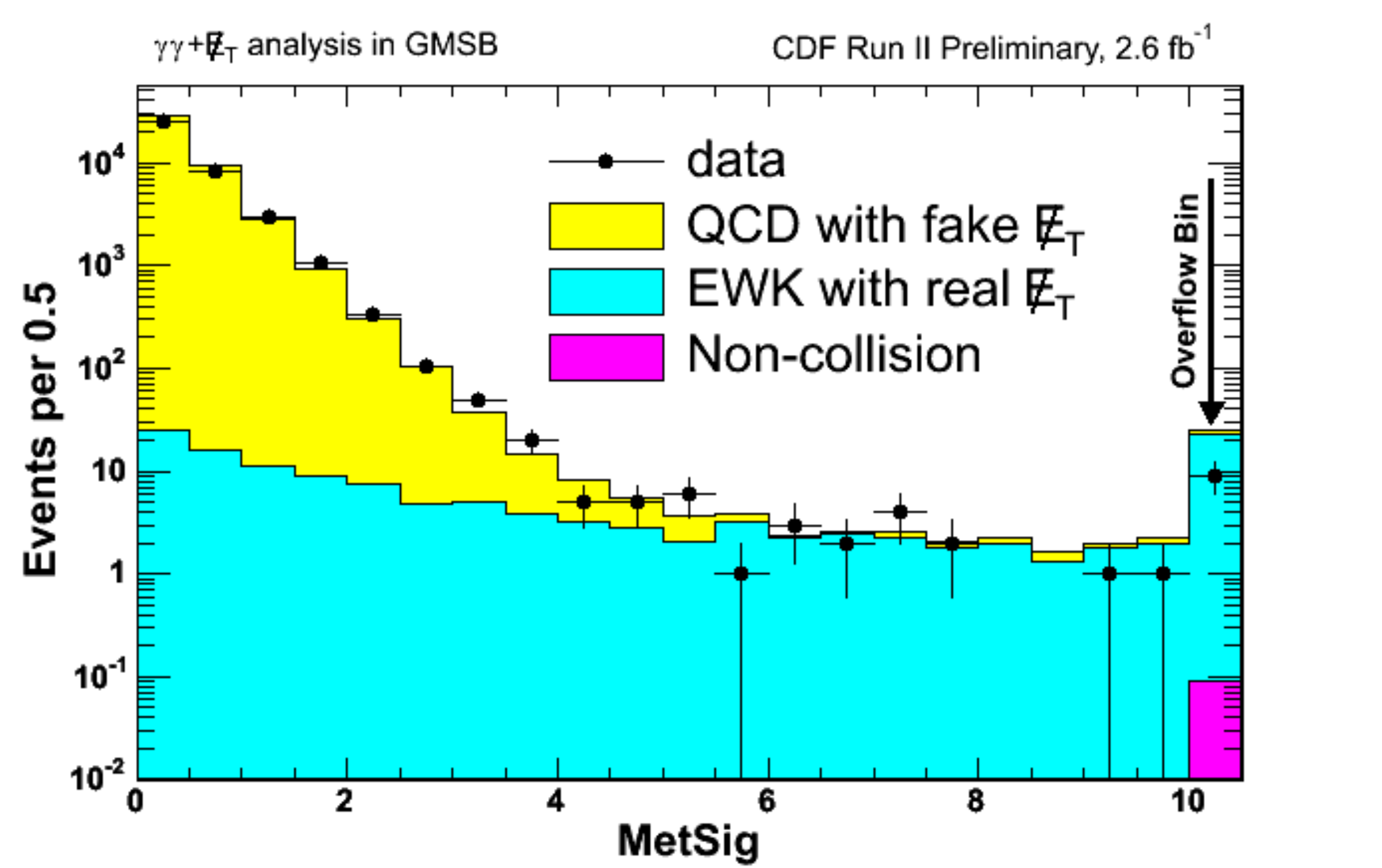}}
\mbox{\includegraphics[width=7.3cm,clip=]{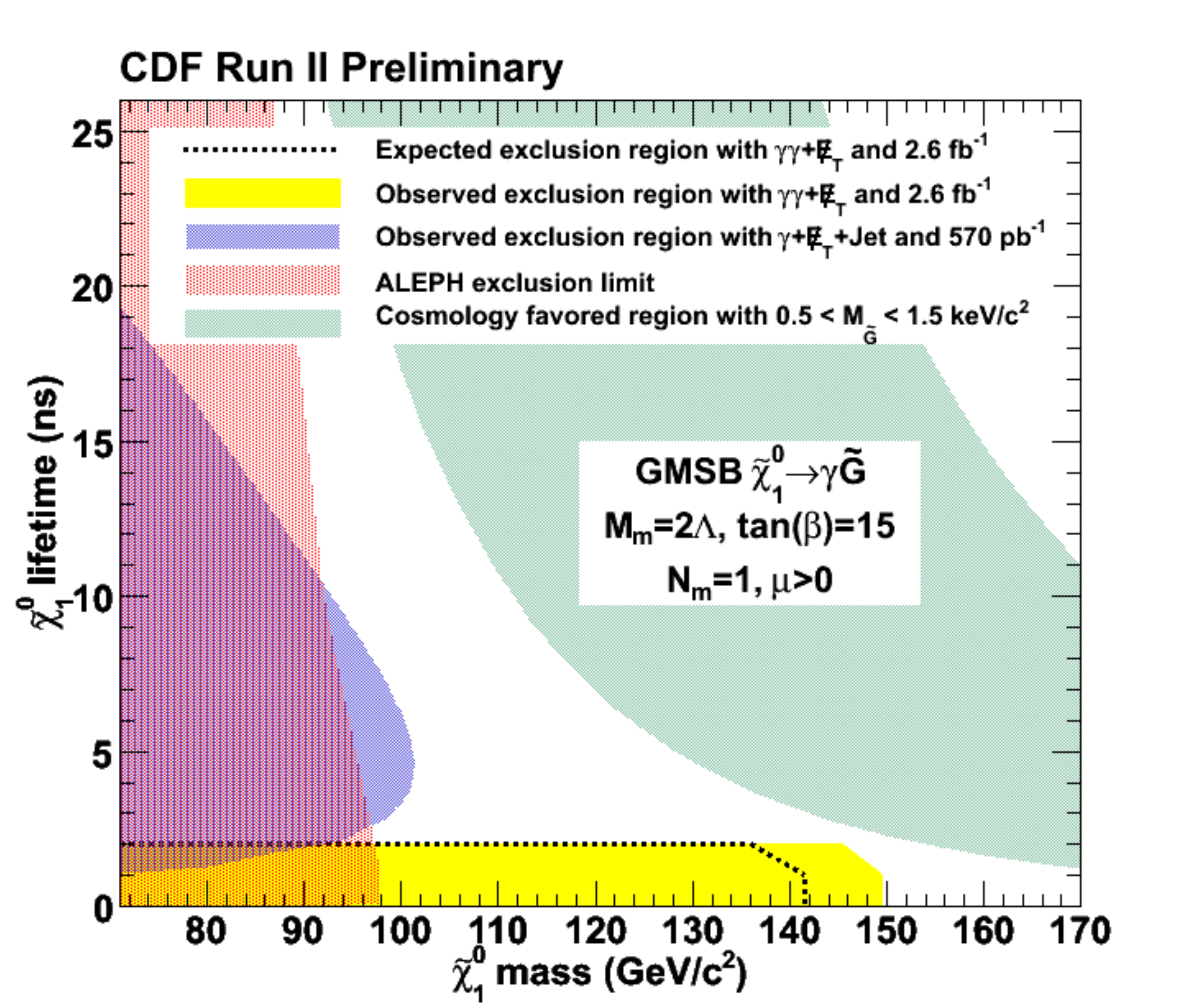}}
\caption{
The $\met$ significance, showing the separation power for mismeasured QCD
and electroweak sources of \met (left), and the limits in the \mneutralino and \neutralino lifetime plane (right).
\label{fig:ggmet}}
\end{figure}

\section{Search for Hidden Valley Dark Photons}
Recently, a subset of HV models with SUSY have emerged which may
simultaneously explain anomalies of the observed spectra of high-energy
cosmic ray positrons and electrons, and provide a solution to the origin of
Dark Matter.  They propose a scenario where dark matter can annihilate to
pairs of so-called dark photons, which then would potentially themselves
decay to pairs of $e^+e^-$ or $\mu^+\mu^-$ through their mixing with the
SM photon.  At the Tevatron, gaugino production could lead to
cascade decays of charginos and neutralinos which could include decays to
both SM photons and these dark photons, whose favored mass range
is on the order of 1 \gev.  The signature, then, will be a SM
photon, two opposite charge leptons spatially close to each other in the
detector, and
\met from the escaping LSP.  \dzero has searched for this signature with 4.1\ifb
of Run II data\cite{Abazov:2009hn}.

Events are selected by requiring a single photon and two leptons.  The
proximity of the leptons to each other requires a careful treatment of the
isolation requirement placed on the leptons.  The main background is QCD
with photon conversions producing pairs of leptons.  After selection, the
dilepton invariant mass is used as a signal discriminant, where the signal
would appear as an excess at the dark photon mass.  No significant excess is
observed, and exclusion limits are set on the lightest chargino mass as a
function of the dark photon mass, shown in Figure~\ref{fig:dphoton}.

\begin{figure}[tb]
\centering
\mbox{\includegraphics[width=7.3cm,clip=]{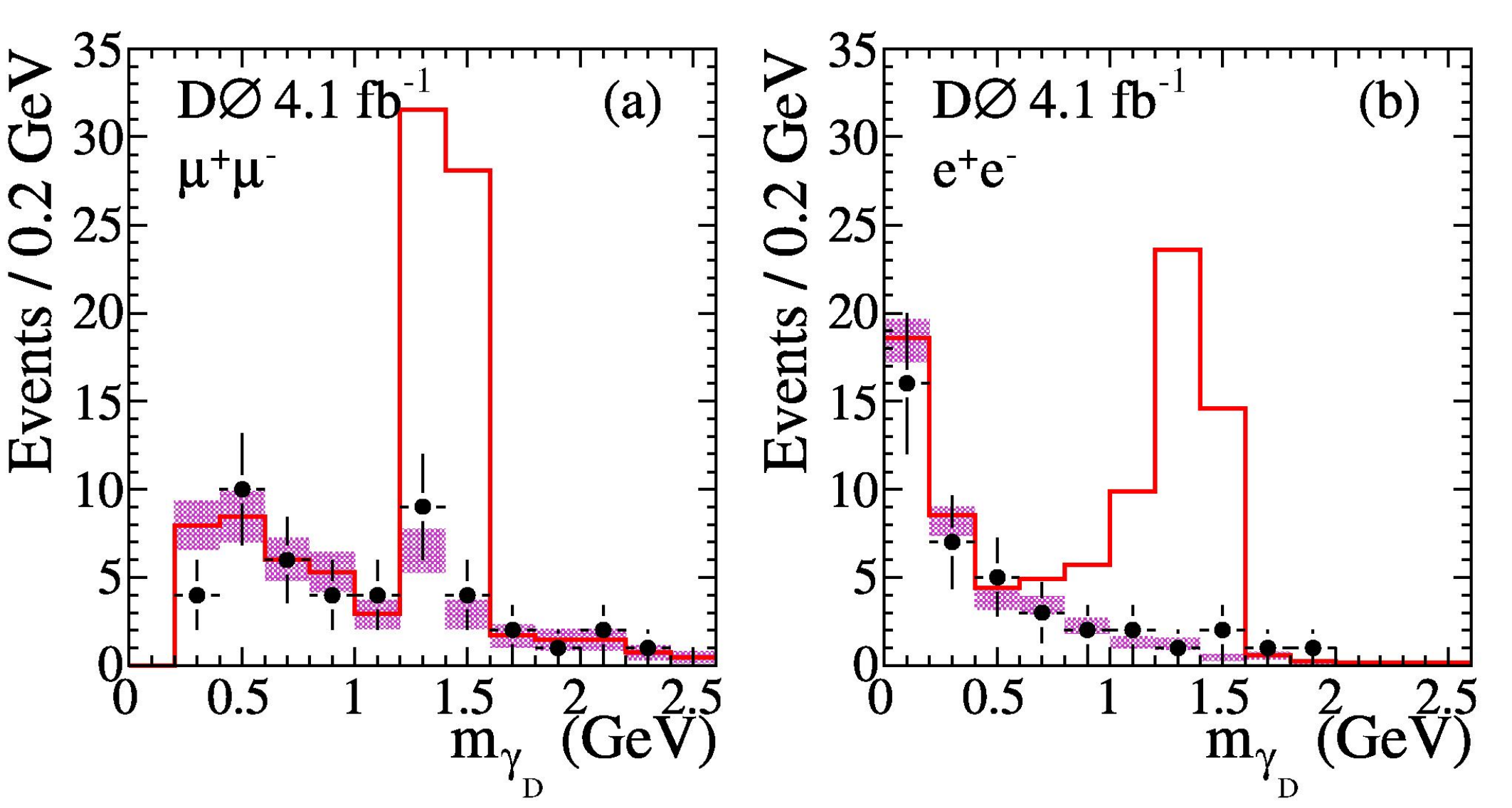}}
\mbox{\includegraphics[width=7.3cm,clip=]{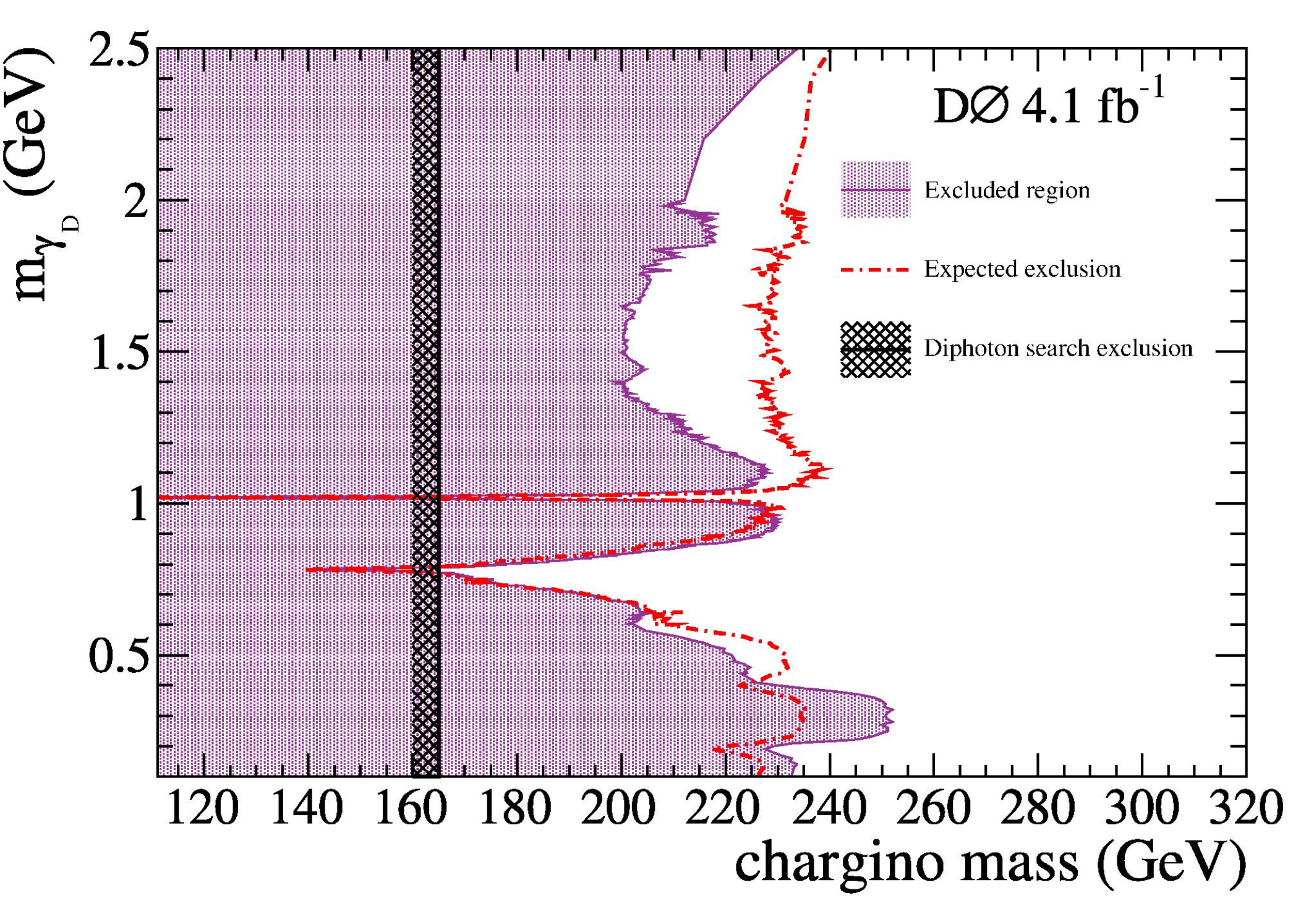}}
\caption{
Dark Photon mass for the muon and electron channels with hypothetical signal shown in solid line(left), and limits in the dark photon mass - chargino mass plane (right).
\label{fig:dphoton}}
\end{figure}

\section{Search for Hidden Valley Long-lived Particles}
Another class of HV models predicts a new, confining gauge group,
weakly coupled to the SM.  Stable configurations of
HV  hadrons, ``v-hadrons'' in this model, can be long-lived.  In
one benchmark model, the SM Higgs boson mixes with the
HV Higgs boson.  In this scenario, the SM Higgs could decay,
through the mixing with the HV Higgs, to pairs of v-hadrons. The
v-hadrons then preferentially couple to the heavier SM quarks due to a
helicity suppression mechanism.  The resulting signature is highly-displaced
secondary vertices with a large number of tracks from the  $b$-quark
decays.  This has been searched for by \dzero with 3.6 \ifb\cite{Abazov:2009ik}.

Events are selected by searching for at least two separated four-track
secondary vertices, with an impact parameter in the transverse plane,
$L_{xy} > 1.6$ cm. The secondary vertex mass and collinearity are used to
select the signal region.  No excess is observed in data.  Assuming the
Higgs always decays to dark-sector v-hadrons, which themselves always decay
to $b\bar{b}$ pairs, limits are set on SM Higgs boson production, shown in Figure~\ref{fig:higgs}.

\begin{figure}[tb]
\centering
\mbox{\includegraphics[width=7.3cm,clip=]{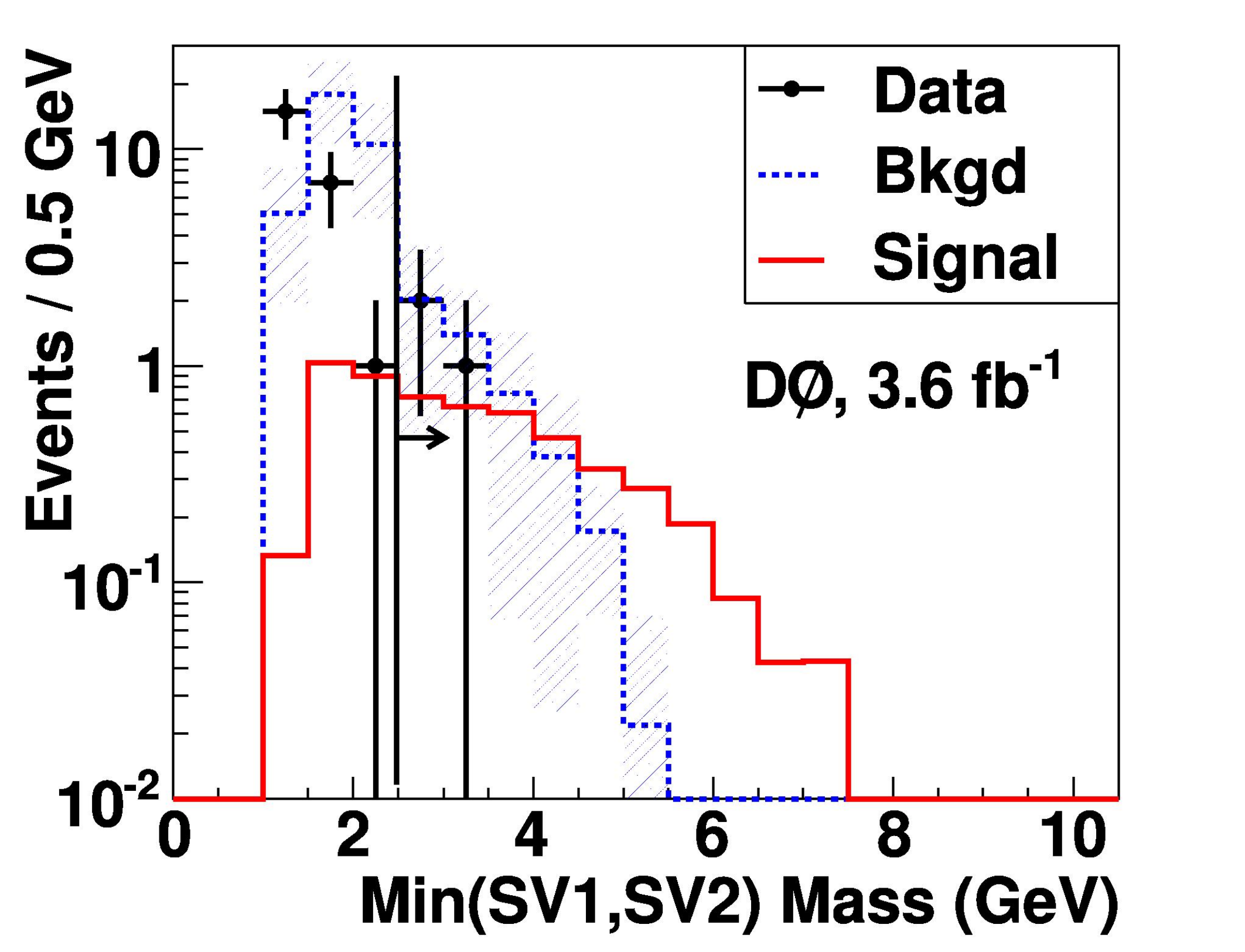}}
\mbox{\includegraphics[width=7.3cm,clip=]{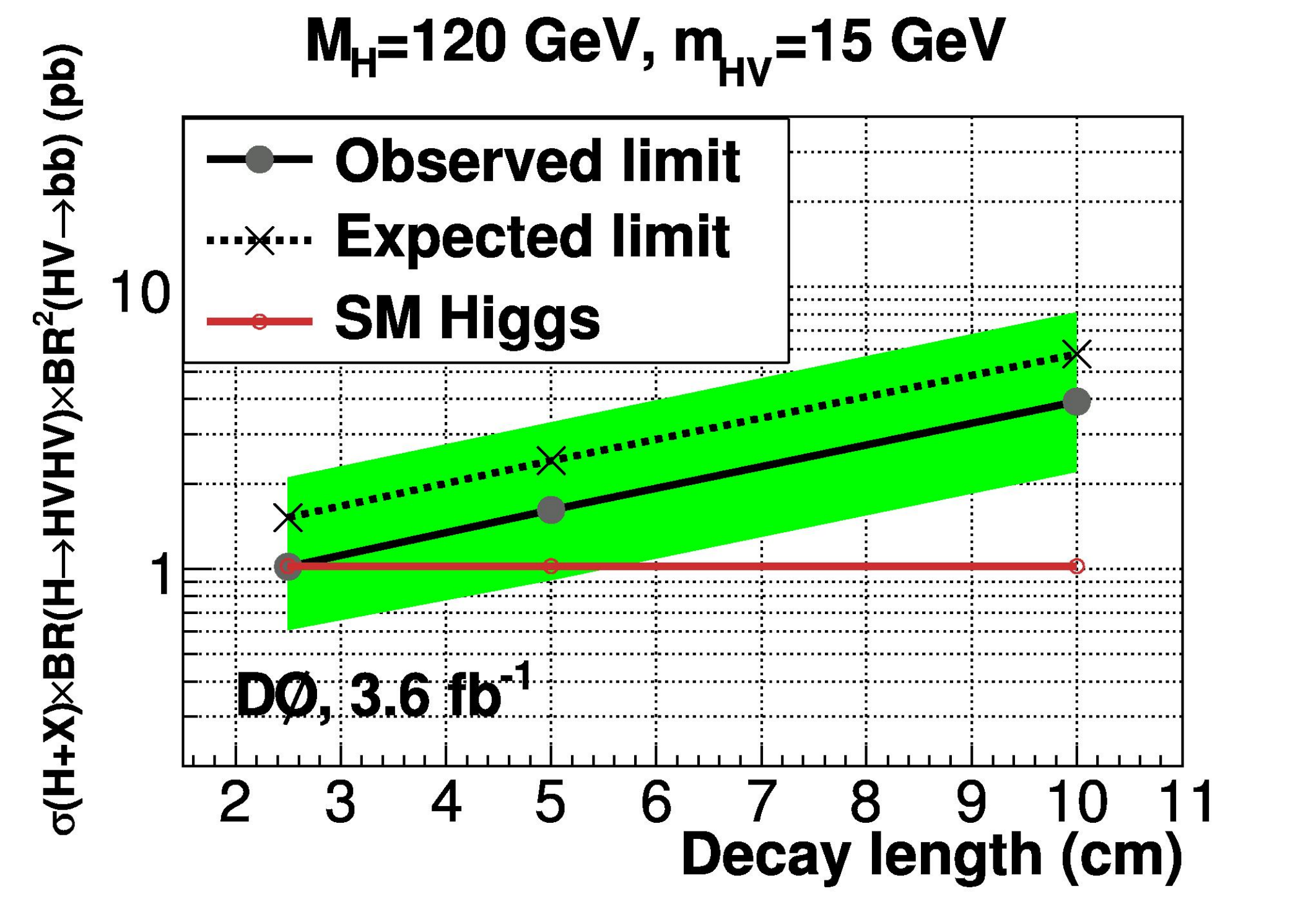}}
\caption{
The smallest secondary vertex mass in the event (left), and limits on
the SM Higgs boson production cross section, for a v-hadron mass of 15 \gevcc (right).
\label{fig:higgs}}
\end{figure}

\section{Conclusions}
The \cdf and \dzero experiments are searching many models and signatures for
evidence of new physics.  None is observed yet, but both experiments have
now recorded more that 8~\ifb, and should have results with this doubling of
the data soon.  Much of the Run II data is yet to be analyzed,
and additional data continues to increase sensitivity to new physics signals.

\section{Acknowledgments}
I would like to thank the conference organizers for the opportunity to
attend the conference.  I also thank my colleagues 
at \dzero and my colleagues at \cdf for help in preparation of this talk.


\section*{References}
\begin{thebibliography}{99}

\bibitem{cdfsbottom}
[arXiv:1005.3600v1 [hep-ex]].

\bibitem{d0sbottom}
[arXiv:1005.2222v2 [hep-ex]].

\bibitem{Aaltonen:2009tp}
  T.~Aaltonen {\it et al.}  [CDF Collaboration],
  ``Search for Supersymmetry with Gauge-Mediated Breaking in Diphoton Events
  with Missing Transverse Energy at CDF II,''
  Phys.\ Rev.\ Lett.\  {\bf 104}, 011801 (2010)
  [arXiv:0910.3606 [hep-ex]].

\bibitem{Abazov:2009hn}
  V.~M.~Abazov {\it et al.}  [D0 Collaboration],
  ``Search for dark photons from supersymmetric hidden valleys,''
  Phys.\ Rev.\ Lett.\  {\bf 103}, 081802 (2009)
  [arXiv:0905.1478 [hep-ex]].


\bibitem{Abazov:2009ik}
  V.~M.~Abazov {\it et al.}  [D0 Collaboration],
  ``Search for Resonant Pair Production of long-lived particles decaying to $b
  \bar{b}$ in $p \bar{p}$ collisions at $\sqrt{s}= 1.96$-TeV,''
  Phys.\ Rev.\ Lett.\  {\bf 103}, 071801 (2009)
  [arXiv:0906.1787 [hep-ex]].


\end{thebibliography}
\end{document}



